# CONFINEMENT AND VISCOELASTIC EFFECTS ON CHAIN CLOSURE DYNAMICS

**Pinaki Bhattacharyya, Rati Sharma and Binny J. Cherayil,[#]**

Dept. of Inorganic and Physical Chemistry, Indian Institute of Science,

Bangalore-560012, INDIA

## ABSTRACT

Chemical reactions inside cells are typically subject to the effects both of the cell's confining surfaces and of the viscoelastic behavior of its contents. In this paper, we show how the outcome of one particular reaction of relevance to cellular biochemistry – the diffusion-limited cyclization of long chain polymers – is influenced by such confinement and crowding effects. More specifically, starting from the Rouse model of polymer dynamics, and invoking the Wilemski-Fixman approximation, we determine the scaling relationship between the mean closure time $t_c$ of a flexible chain (no excluded volume or hydrodynamic interactions) and the length $N$ of its contour under the following separate conditions: (a) confinement of the chain to a sphere of radius $D$, and (b) modulation of its dynamics by colored Gaussian noise. Among other results, we find that in case (a) when $D$ is much smaller than the size of the chain, $t_c \sim ND^2$, and that in case (b), $t_c \sim N^{2/(2-2H)}$, $H$ being a number between 1/2 and 1 that characterizes the decay of the noise correlations. $H$ is not known à priori, but values of about 0.7 have been used in the successful characterization of protein conformational dynamics. At this value of $H$ (selected for purposes of illustration), $t_c \sim N^{3.4}$, the high scaling exponent reflecting the slow relaxation of the chain in a viscoelastic medium.

[#] Corresponding author. Email: cherayil@ipc.iisc.ernet.in

# 1. INTRODUCTION

Despite their intrinsic stochasticity, molecular processes inside cells have somehow evolved to produce reaction pathways that seem exquisitely choreographed. How this is achieved in spaces that are typically only a few micrometers across and that are extremely congested is one of many unanswered questions that presently defines the limits of our understanding of living matter. We believe that a deeper appreciation of the effects of confinement and crowding on single-molecule reaction dynamics would go some way towards advancing our knowledge of life's operations at the microscopic level. To this end, we attempt to show, in this paper, how such factors can influence the outcome of one particular reaction of considerable biological significance: chain cyclization.

The cyclization of biopolymers like DNA and RNA is significant for a number of reasons; it facilitates the interaction of DNA-bound proteins with distant target sites, thereby regulating gene expression;[1] it leads to the formation of compact tertiary structures, which may be important in the packaging of genetic information;[2] and it likely plays a role in mediating the long-range interactions that signal the start of replication at different locations on the polynucleotide backbone.[3] The importance of the cyclization reaction is, of course, not confined to biopolymers alone, but extends to other polymeric systems as well, including many with industrial and commercial applications.[4]

Over the last several years, a great deal of theoretical research has therefore been devoted to the development of models of the dynamics of chain cyclization,[5-19] with a special emphasis on understanding how the mean reaction time varies with chain properties, particularly the molecular weight. Because of the intrinsic many-body character of the cyclization reaction in polymers (which in any realistic description requires a treatment of all the monomers in the chain, including their mutual excluded volume and hydrodynamic interactions), calculations of reaction times and similar quantities are generally non-trivial, and require numerous approximations. They are rendered even more non-trivial when boundaries are present, or when the medium is viscoelastic, and to the best of our knowledge, they have so far not been attempted at all under these conditions. But it is precisely these conditions that are generally encountered by long chain molecules inside cells, microfluidic devices, or other confined spaces. Fully extended DNA, for instance, can exceed by several orders of magnitude the size of the region to which it is typically confined; moreover, its dynamics inside the cell is mediated by a fluid that likely contains a dense mixture of entangled polymeric chains, rendering its surroundings highly viscoelastic. In these kinds of surroundings, which may also be met by polymers in non-biological contexts, molecular diffusion is often anomalous[20] (meaning the center of mass motion varies sub-linearly with time), and this circumstance is likely to affect the cyclization reaction.

To what extent such circumstances – confinement and crowding, specifically – affect the chain length dependence of the mean reaction time is the question we shall attempt to address in this paper. The confinement problem is considered by studying the cyclization reaction using a Rouse chain that moves inside a sphere. The crowding problem is considered by studying the reaction using an unconfined "renormalized" Rouse chain that moves through a complex fluid. In both studies, the actual calculations are carried out within the framework of an approach developed by Wilemski and Fixman[5], and since used, in an important paper by Pastor, Szabo and Zwanzig[10], to determine the mean cyclization time of a polymer at the theta point as a function of various chain properties. The WF formalism has also been used to calculate the mean closure times for unconfined polymers with bending stiffness[13] and with long-range excluded volume interactions,[17] and is generally felt to provide a sensible approach to the study of cyclization dynamics.[6,7,21] However, its utility in analyzing actual data may depend on how closely experimental conditions ensure that the closure reaction is diffusion limited. Experiments performed on synthetic DNA and RNA, for instance, often rely on fluorescence quenching to measure reaction times, and in that case the kinetics of electron transfer, which introduce other timescales into the problem, may need to be accounted for. The WF approximation may then no longer be entirely trustworthy.[8,14,19] It is therefore only when quenching is very efficient and its influence effectively negligible that the WF approximation can be assumed to hold good. In using this approximation in our calculations, therefore, we are assuming implicitly that diffusion-limited conditions do in fact prevail. .

The next section is a brief review of the WF formalism, highlighting the role of the time correlation function of the end-to-end distance in the evaluation of the mean closure time $t_c$. Section 3 discusses how this time correlation function is itself calculated for confined polymers and polymers in complex fluids. And finally, sec. 4 discusses the results of these calculations, and their implications.

## 2. THEORETICAL BACKGROUND

Consider, to begin with, a polymer of $n$ monomers with reactive ends in an *unbounded* viscous fluid at temperature $T$. The configuration of the polymer at time $t$ is specified by the set of monomer positions $\{\mathbf{r}\} \equiv (\mathbf{r}_1, \mathbf{r}_2, \ldots, \mathbf{r}_n)$. If $\psi(\{\mathbf{r}\}, t)$ denotes the probability density that the first monomer is between $\mathbf{r}_1$ and $\mathbf{r}_1 + d\mathbf{r}_1$, the second between $\mathbf{r}_2$ and $\mathbf{r}_2 + d\mathbf{r}_2$, and so on, the evolution of $\psi(\{\mathbf{r}\}, t)$ is given by the following equation

$$\frac{\partial \psi(\{\mathbf{r}\}, t)}{\partial t} = \mathsf{D}\psi(\{\mathbf{r}\}, t) - \lambda S(\{\mathbf{r}\})\psi(\{\mathbf{r}\}, t) \tag{1}$$

Here $S(\{\mathbf{r}\})$ is a sink function (to be specified later), $\lambda$ is a reaction rate, and $\mathsf{D} \equiv D_0 \sum_{i=1}^{n} \left[ \nabla_i^2 + \nabla_i \cdot \nabla_i \beta U \right]$ is a generalized diffusion operator in which $\nabla_i \equiv \partial/\partial \mathbf{r}_i$, $D_0 \equiv k_B T / \zeta$ is a diffusion coefficient, with $\zeta$ a monomer friction coefficient and $k_B T \equiv 1/\beta$ the Boltzmann factor, and $U = U(\{\mathbf{r}\})$ is the intermolecular potential.

As has been shown elsewhere[5,10], for a chain whose dynamics are governed by Eq. (1) and for which $S(\{\mathbf{r}\})$ is a function solely of the distance between one end of the chain and the other, the mean cyclization time $t_c$ is given by

$$t_c = \int_0^\infty dt \left( \frac{C(t)}{C(\infty)} - 1 \right) \tag{2}$$

where

$$C(t) = \int d\mathbf{R} \int d\mathbf{R}_0 S(\mathbf{R}) G(\mathbf{R}, t | \mathbf{R}_0, t) S(\mathbf{R}_0) \psi_{eq}(\mathbf{R}_0), \tag{3}$$

$\mathbf{R}$ and $\mathbf{R}_0$ are, respectively, the end-to-end distances of the chain at time $t$ and time 0, $G(\mathbf{R}, t | \mathbf{R}_0, t)$ is the conditional probability density that a chain with the end-to-end distance $\mathbf{R}_0$ at 0 has the end-to-end distance $\mathbf{R}$ at time $t$, and $\psi_{eq}(\mathbf{R}_0)$ is the initial equilibrium distribution of end-to-end distances. This expression has been arrived at on the basis of the closure scheme introduced by Wilemski and Fixman[5] in which the exact solution to Eq. (1) is replaced by an approximate expression involving the product of the equilibrium chain distribution and a self-consistently determined time-dependent correction. It has been further assumed that $\lambda \to \infty$, which implies that the ends of the chain react instantaneously and irreversibly whenever they satisfy the distance constraint imposed by the sink function.

The function $G(\mathbf{R},t|\mathbf{R}_0,t)$ is central to the calculation of $t_c$, but it is not known in general. However, if $\mathbf{R}$ is a Gaussian stochastic process, it is given explicitly by[22]

$$G(\mathbf{R},t|\mathbf{R}_0,0) = \left(\frac{3}{2\pi\langle\mathbf{R}^2\rangle_{eq}}\right)^{3/2} \frac{1}{[1-\varphi^2(t)]^{3/2}} \exp\left[-\frac{3(\mathbf{R}-\varphi(t)\mathbf{R}_0)^2}{2\langle\mathbf{R}^2\rangle_{eq}(1-\varphi^2(t))}\right] \quad (4)$$

where $\varphi(t) \equiv \langle\mathbf{R}(t)\cdot\mathbf{R}(0)\rangle/\langle\mathbf{R}^2\rangle_{eq}$ is the normalized time correlation function of the end-to-end distance, the angular brackets denoting an ensemble average over chain configurations. In the present model of chain dynamics, excluded volume and hydrodynamic interactions are assumed to be absent, and $\mathbf{R}$ is therefore a sum of a large number of bonds whose random orientations are driven by white noise (or colored Gaussian noise in the case of diffusion through a viscoelastic medium.) $\mathbf{R}$ itself is therefore a Gaussian random variable, and Eq. (4) is a satisfactory description of its time-dependent conditional probability density.

Making the further assumption that $S(\mathbf{R})$ depends only on $R=|\mathbf{R}|$, one can carry out the angular integrations in Eq. (3) analytically,[23] with the result that

$$C(t) = 16\pi^2 \left(\frac{3}{2\pi\langle R^2\rangle_{eq}}\right)^3 \frac{1}{[1-\varphi^2(t)]^{3/2}} \int_0^\infty dR R^2 S(R) \int_0^\infty dR_0 R_0^2 S(R_0)$$

$$\times \frac{\langle R^2\rangle_{eq}(1-\varphi^2(t))}{3\varphi(t)RR_0} \sinh\left(\frac{3\varphi(t)RR_0}{\langle R^2\rangle_{eq}(1-\varphi^2(t))}\right) \exp\left(-\frac{3(R^2+R_0^2)}{2\langle R^2\rangle_{eq}(1-\varphi^2(t))}\right) \quad (5)$$

Further simplification is possible if the sink function is chosen to correspond to the delta function, i.e., $S(R) = \delta(R-a)$. It then follows that[10]

$$\frac{C(t)}{C(\infty)} = \frac{1}{2x_0\varphi(t)\sqrt{1-\varphi^2(t)}} \sinh\left(\frac{2x_0\varphi(t)}{1-\varphi^2(t)}\right) \exp\left(-\frac{2x_0\varphi^2(t)}{1-\varphi^2(t)}\right) \quad (6)$$

where $x_0 = 3a^2/2\langle R^2\rangle_{eq}$. This expression is singular at $t=0$ where $\varphi(t)=1$, but it can be rewritten in the more convenient form

$$\frac{C(t)}{C(\infty)} \approx \left(1-\varphi^2(t) + \frac{4}{3}x_0\varphi^2(t) + \cdots\right)^{-3/2} \quad (7)$$

following the approach adopted by Pastor et al.[10] whereby Eq. (6) is expanded in powers of $x_0$ and then re-expressed in a resummed form that to the same order in $x_0$ is identical to the original expansion. The above expression proves to be convenient for deriving analytic scaling results for $t_c$.

The next two sections describe the treatment of polymer cyclization rates when the chain is confined to a sphere and when it moves through a viscoelastic continuum.

## 3. CHAIN CYCLIZATION DYNAMICS IN A SPHERE

### A. Effective Rouse description

To explore the influence of boundaries on chain cyclization rates we calculate, in this section, the mean first passage cyclization time of a simple Gaussian chain confined to a spherical cavity[24] of radius $D$, with one end fixed at the center of the sphere. (No real loss of generality is entailed by imposing this constraint, but it does simplify the algebra.) Even this seemingly simple system presents a number of analytical challenges that can only be met by the introduction of several further approximations (going beyond those defined by the Wilemski-Fixman[5] closure scheme.) Chief among these is an approximation for the treatment of polymer dynamics in the presence of boundaries. Although the equilibrium statistical mechanics of polymer-surface interactions is well developed,[25] extensions of the methodology to the dynamical regime are in general non-trivial. But a relatively simple approach to the problem, which is nevertheless systematic and well-controlled, has recently been introduced by Kalb and Chakraborty.[26] It is based on the idea that the confined chain admits of a representation in terms of decoupled Rouse modes in which the geometry of the external surface is wholly incorporated into the relaxation times of the internal modes. These relaxation times can be expressed in terms of the *equilibrium* conformation of the chain under confinement, and can therefore be calculated rigorously.

To implement this approach, we recall, first of all, that for a free Rouse chain, the equation of motion of individual monomers (in continuum notation) is given by[27]

$$\zeta \frac{\partial \mathbf{r}(\tau,t)}{\partial t} = k \frac{\partial^2 \mathbf{r}(\tau,t)}{\partial \tau^2} + \mathbf{\theta}(\tau,t) \qquad (8)$$

where $\mathbf{r}(\tau,t)$ is the spatial location, at time $t$, of the monomer at the point $\tau$ on the chain backbone, $k \equiv 3k_B T / b^2$, $b$ being the Kuhn length of the chain, and $\mathbf{\theta}(\tau,t)$ is a random force (acting on the monomer at $\tau$) whose statistical properties are entirely defined by the following correlations: $\langle \theta_\alpha(\tau,t) \rangle = 0$ and $\langle \theta_\alpha(\tau,t)\theta_\beta(\tau',t') \rangle = 2\zeta k_B T \delta_{\alpha\beta} \delta(t-t')\delta(\tau-\tau')$. Equation (8) can be solved by expanding the variables $\mathbf{r}(\tau,t)$ in a set of independent normal modes $\mathbf{X}_p(t)$, using the relations

$$\mathbf{X}_p(t) = N^{-1} \int_0^N d\tau \cos(p\pi\tau/N)\mathbf{r}(\tau,t), \ p \geq 0 \text{ and } \mathbf{r}(\tau,t) = \mathbf{X}_0(t) + 2\sum_{p=1}^\infty \mathbf{X}_p(t)\cos(p\pi\tau/N),$$

$N$ being the contour length of the chain. When this is done, Eq. (8) is transformed to the decoupled equations

$$\zeta_p \frac{\partial \mathbf{X}_p(t)}{\partial t} = -k_p \mathbf{X}_p(t) + \mathbf{f}_p(t) \tag{9}$$

where $\zeta_0 = N\zeta$, $\zeta_p = 2N\zeta$ for $p \geq 1$, $k_p = 2\pi^2 k p^2 / N$ for $p \geq 0$, and $f_p(t) = 2\int_0^N d\tau \mathbf{\theta}(\tau,t)\cos(p\pi\tau/N)$, with $\langle \mathbf{f}_p(t) \rangle = 0$ and $\langle f_{p\alpha}(t)f_{q\beta}(t') \rangle = 2\zeta_p k_B T \delta_{pq} \delta_{\alpha\beta} \delta(t-t')$. Equation (9) is easily solved in closed form, the solution being $\mathbf{X}_p(t) = \mathbf{X}_p(0)e^{-k_p t/\zeta_p} + \zeta_p^{-1} \int_0^t dt' e^{-k_p(t-t')/\zeta_p} \mathbf{f}_p(t')$.

In terms of the above normal modes, the end-to-end vector $\mathbf{R}(t)$ is given by $\mathbf{R}(t) = -4\sum_{p \text{ odd}} \mathbf{X}_p(t)$, so the end-to-end correlation function is

$$\langle \mathbf{R}(t)\cdot\mathbf{R}(0) \rangle = 16 \sum_{p \text{ odd}} \langle \mathbf{X}_p(t)\cdot\mathbf{X}_p(0) \rangle = 16 \sum_{p \text{ odd}} \frac{3k_B T}{k_p} \exp(-t/\tau_p) \tag{10}$$

where the second equality follows from the solution of Eq. (9) and the properties of $\mathbf{f}_p(t)$. The parameter $\tau_p$ is the characteristic relaxation time of the mode $p$, which is given by $\tau_p = \tau_1/p^2$, $\tau_1$ being the longest relaxation time, which itself is given by $\tau_1 = \zeta_1/k_1$; $k_1$, in turn, is given by the equilibrium average $k_1 = 3k_BT/\langle \mathbf{X}_1^2 \rangle_{eq}$.

Since Eq. (10) is a sum of rapidly decreasing exponentials, the time correlation function of the end-to-end vector is governed mainly by the first mode. To a reasonable approximation, therefore,

$$\langle \mathbf{R}(t) \cdot \mathbf{R}(0) \rangle \approx 16 \langle \mathbf{X}_1^2 \rangle_{eq} \exp(-k_1 t/\zeta_1) \tag{11}$$

and so $\langle \mathbf{R}^2(0) \rangle \approx 16 \langle \mathbf{X}_1^2 \rangle_{eq}$. Thus, if the chain is initially in equilibrium (and this assumption is necessary in the implementation of the Wilemski-Fixman method[5]), this means that

$$\langle \mathbf{R}(t) \cdot \mathbf{R}(0) \rangle \approx \langle \mathbf{R}^2 \rangle_{eq} \exp\left(-24 k_B T t / N \zeta \langle \mathbf{R}^2 \rangle_{eq}\right) \tag{12}$$

In the spirit of approximations that treat the dynamics of chains with excluded volume, hydrodynamic or monomer-surface interactions by means of a Rouse model with effective or scaled parameters,[26,27] we now assume that under confinement $\langle \mathbf{R}(t) \cdot \mathbf{R}(0) \rangle$ has exactly the structure of Eq. (12), and that all of the effects of confinement are contained in the quantity $\langle \mathbf{R}^2 \rangle_{eq}$, which is amenable to calculation by standard *equilibrium* statistical mechanical methods. That calculation is described in the next section, but before discussing its details, it is useful to note that the approximations of Eqs. (10) – (12) reduce the function $\varphi(t)$ that is needed in the evaluation of the cyclization time to the expression

$$\varphi(t) = e^{-\alpha t} \tag{13}$$

where $\alpha = 24 k_B T / N\zeta \langle \mathbf{R}^2 \rangle_{eq}$; Eq. (13) will be recognized as the defining relation of the so-called harmonic spring model,[6] about which more will be said later.

**B. Equilibrium dimensions of a spherically confined Gaussian polymer**

The mean square end-to-end distance $\langle \mathbf{R}^2 \rangle_{eq}$ of a free Gaussian polymer of contour length $N$, one end of which is located at the origin, is given in general by the relation

$$\langle \mathbf{R}^2 \rangle_{eq} = \frac{1}{Z_0} \int d\mathbf{R}\, \mathbf{R}^2 G_0(\mathbf{R}, N) \tag{14}$$

where $Z_0 = \int d\mathbf{R}\, G_0(\mathbf{R}, N)$, and $G_0(\mathbf{R}, N)$ is the chain propagator, defined formally by the path integral

$$G_0(\mathbf{R}, N) = \int_{\mathbf{r}(0)=\mathbf{0}}^{\mathbf{r}(N)=\mathbf{R}} D[\mathbf{r}]\exp\left[-\frac{3}{2l^2}\int_0^N d\tau \left(\frac{\partial \mathbf{r}(\tau)}{\partial \tau}\right)^2\right] \tag{15}$$

which may be shown to satisfy the differential equation[28]

$$\left(\frac{\partial}{\partial N} - \frac{b^2}{6}\nabla_{\mathbf{R}}^2\right) G_0(\mathbf{R}, N) = \delta(N)\delta(\mathbf{R}) \tag{16}$$

whose solution is $G_0(\mathbf{R}, N) = (3/2\pi N b^2)^{3/2} \exp(-3\mathbf{R}^2/2Nb^2)$. When the polymer is confined to a certain volume by a surface, such as the surface of a sphere, the corresponding propagator, $G(\mathbf{R}, N)$, may still be derived from Eq. (16), but it must now satisfy the boundary conditions appropriate to the nature of the confining geometry. As shown in Appendix A, if this geometry is that of a sphere of radius $D$ (requiring $G(\mathbf{R}, N)$ to vanish at $R \equiv |\mathbf{R}| = D$), $G(\mathbf{R}, N)$ is given by[24]

$$G(R,N) = \frac{1}{2^{3/2}\pi D^{5/2} R^{1/2}} \sum_{n=1}^{\infty} \frac{1}{J_{3/2}^2(y_{0n})} \sqrt{y_{0n}} \, J_{1/2}(y_{0n} R/D) \exp\left[-y_{0n}^2 Nb^2/6D^2\right] \quad (17)$$

where $J_\nu(x)$ is the Bessel function of order $\nu$, and $y_{0n}$ is the $n$th zero of the Bessel function of order 1/2, i.e., $J_{1/2}(y_{0n}) = 0$.

The calculation of $\langle R^2 \rangle_{eq}$ using the analogue of Eq. (14), viz., $\langle R^2 \rangle_{eq} = Z^{-1} \int d\mathbf{R} R^2 G(\mathbf{R}, N)$, where $Z = \int d\mathbf{R} G(\mathbf{R}, N)$ and $\int d\mathbf{R} = \int_0^{2\pi} d\phi \int_0^{\pi} d\theta \sin\theta \int_0^D dR R^2$ presents no special difficulties, though it is necessary to refer to tabulated results[29] for values of integrals involving the Bessel functions. In this way, we find that

$$\langle R^2 \rangle_{eq} = D^2 \left[ 1 - \frac{6\sum_{n=1}^{\infty}(-1)^{n+1} n^{-2} \exp(-n^2 \pi^2 \Lambda)}{\pi^2 \sum_{n=1}^{\infty}(-1)^{n+1} \exp(-n^2 \pi^2 \Lambda)} \right] \quad (18)$$

where $\Lambda \equiv Nb^2/6D^2$. One may verify that in the limit $D \gg 1$, $\langle R^2 \rangle_{eq} \to 6D^2 \Lambda = Nb^2$, and that in the limit $D \ll 1$, $\langle R^2 \rangle_{eq} \to D^2(1 - 6/\pi^2)$, independent of $N$.

This result, Eq. (18), together with Eq. (13) provide the ingredients for the determination of the cyclization time. Specifically, by substituting Eqs. (7) and (13) into (2), we find that

$$t_c = \frac{N\zeta \langle R^2 \rangle_{eq}}{48 k_B T} Z_1(x_0) \quad (19a)$$

where

$$Z_1(x_0) = \int_0^1 dy \frac{1}{y}\left(\frac{1}{(1-\sigma y)^{3/2}} - 1\right) \qquad (19b)$$

with $\sigma = 1 - 4x_0/3$ and $x_0$, as already defined, given by $x_0 = 3a^2/2\langle \mathbf{R}^2 \rangle_{eq}$, $a$ being the reaction radius. A discussion of these equations will be presented in Sec. 5.

## 4. CHAIN CYCLIZATION DYNAMICS IN A CROWDED ENVIRONMENT

In this section we turn our attention to the calculation of the mean closure time for a polymer in a fluid that by virtue of a high density of other macromolecular species in it is viscoelastic. The cyclization of a polymer in such fluids (of which the cytoplasm is clearly an example) is a many-chain problem, and is difficult to treat in complete generality. But as shown by Schweizer using a projection operator formalism applied to a polymer melt,[30] the problem can be reduced to one involving just a single tagged chain whose monomer dynamics are described, approximately, by a generalized Langevin equation (GLE).[31] All the effects of the other chains in the medium are then contained in the memory function of this GLE, which in Schweizer's calculations was the object of primary interest, and the focus of efforts to develop theoretical models of. In the present calculations (as in some earlier ones[32]), we take the GLE as the point of departure for exploring chain cyclization times in crowded environments, but no longer regard the memory function as a quantity to be determined rigorously from first principles; instead, we fix its functional form *at the outset* by assuming – on the basis of earlier corroborative evidence from experiment,[33] theory[34] and numerical simulations[35] – that the random thermal forces that govern chain dynamics in a viscoelastic medium can be described by the stochastic process known as fractional Gaussian noise (fGn).[36] Under this assumption, the memory function becomes a simple power law in time, and the resulting GLE then becomes amenable to analytic treatment, as we now show.

The general structure of the GLE obtained from Schweizer's projection operator approach is, in the continuum notation of the previous section,

$$\zeta \int_0^N d\tau' \int_0^t dt' \Gamma(|\tau-\tau'|, t-t') \frac{\partial \mathbf{r}(\tau', t')}{\partial t'} = k \frac{\partial^2 \mathbf{r}(\tau, t)}{\partial \tau^2} + \mathbf{F}(\tau, t) \qquad (20)$$

Here $\mathbf{F}(\tau,t)$ is the random thermal referred to above, and at time $t$ it acts on the monomer at the point $\tau$; $\Gamma(|\tau-\tau'|, t-t')$ is the memory function, which is related to $\mathbf{F}(\tau,t)$ by a fluctuation-dissipation theorem:[31] $\langle F_\alpha(\tau,t) F_\beta(\tau',t') \rangle = \zeta k_B T \delta_{\alpha\beta} \Gamma(|\tau-\tau'|, t-t')$; and $k \equiv 3k_B T/b^2$ is the spring constant introduced earlier. In deriving this equation, it has been assumed that inertial contributions are negligible, and that hydrodynamic interactions are screened out. By choosing $\mathbf{F}(\tau,t)$ to correspond to fGn, we see from the fluctuation-dissipation relation that $\Gamma \propto |t-t'|^{2H-2}$, where $H$, the Hurst index, is a number between 1/2 and 1 that characterizes the degree of correlation between force fluctuations at different instants of time.

Equation (20) may be solved by the same normal mode method that was used in the analysis of the Rouse model. That method yields the decoupled equation:

$$\zeta_p \int_0^t dt' \Gamma_p(t-t') \frac{\partial \mathbf{X}_p(t')}{\partial t'} = -\frac{k_p}{N} \mathbf{X}_p(t) + \mathbf{F}_p(t) \qquad (21)$$

where $\Gamma_p(t-t') = N^{-1} \int_0^N d\tau \Gamma(\tau, t-t') \cos(p\pi\tau/N)$ and $\mathbf{F}_p(t) = 2N^{-1} \int_0^N d\tau \mathbf{F}(\tau,t) \cos(p\pi\tau/N)$. From the properties of $\mathbf{F}(\tau,t)$, $\mathbf{F}_p(t)$ is characterized by these statistical correlations: $\langle \mathbf{F}_p(t) \rangle = 0$ and $\langle F_{p\alpha}(t) F_{q\beta}(t') \rangle = \zeta_p k_B T N^{-1} \delta_{\alpha\beta} \delta_{pq} \Gamma_p(t-t')$. The identification of $\mathbf{F}(\tau,t)$ with fGn means that $\Gamma_p(t-t') = 2H(2H-1) N^{-1} |t-t'|^{2H-2}$ (assuming no mode dependence of the memory function, an assumption largely justified by Schweizer's calculations[30])

As before, the key ingredient in the calculation of the cyclization time is the time correlation function of the end-to-end distance, which is obtained from the time correlation function of $\mathbf{X}_p(t)$. The latter is easily found from Eq. (21) using Laplace transforms. The result is

$$\langle \mathbf{X}_p(t) \cdot \mathbf{X}_p(0) \rangle = \langle \mathbf{X}_p^2(0) \rangle E_{2-2H}\left[ -(t/\tau_{RR})^{2-2H} \right] \tag{22}$$

where $E_a(z) \equiv \sum_{k=0}^{\infty} z^k / \Gamma(ak+1)$ is the Mittag-Leffler function,[37] $\tau_{RR}$ is a characteristic relaxation time given by $\tau_{RR} = \left( \zeta_p \Gamma(2H+1)/k_p \right)^{1/(2-2H)}$, and $\langle \mathbf{X}_p^2(0) \rangle$ is $3k_B T/k_p$ as before. From these results, using Eq. (10), we find that

$$\varphi(t) \equiv \frac{1}{\langle \mathbf{R}^2(0) \rangle} \langle \mathbf{R}(t) \cdot \mathbf{R}(0) \rangle = \frac{\sum_{p \text{ odd}} k_p^{-1} E_{2-2H}\left[ -(t/\tau_{RR})^{2-2H} \right]}{\sum_{p \text{ odd}} k_p^{-1}}$$

$$= \frac{8}{\pi^2} \sum_{p=1}^{\infty} \left[ \frac{1}{(2p-1)^2} E_{2-2H}\left( -\frac{(2p-1)^2 k\pi^2 t^{2-2H}}{\zeta N^2 \Gamma(2H+1)} \right) \right] \tag{23}$$

This function is shown in Fig.1. The curve corresponding to $H = 1/2$ describes simple exponential decay; the higher values of $H$ describe chains that are increasingly sluggish, which is the behavior expected in media that are very crowded, such as a polymer melt, or a concentrated polymer solution.

The cyclization time is now given by

$$t_c = \frac{1}{2-2H}\left(\frac{N^2 \zeta \Gamma(2H+1)}{k\pi^2}\right)^{1/(2-2H)} Z_2(x_0) \qquad (24a)$$

where

$$Z_2(x_0) = \int_0^\infty dy\, y^{(2H-1)/(2-2H)} \left(\frac{1}{(1-\sigma\varphi^2(y))^{3/2}} - 1\right) \qquad (24b)$$

with $\varphi(y) = 8\pi^{-2}\sum_{p=1}^{\infty}(2p-1)^{-2} E_{2-2H}[-(2p-1)^2 y]$.

## 5. RESULTS AND DISCUSSION

The primary aim of these calculations has been to better understand how a reaction relevant to cellular biochemistry – chain cyclization – is affected by steric constraints. To this end, we have sought to determine the scaling relationship between the mean cyclization time $t_c$ of a polymer having no excluded volume or hydrodynamic interactions and its contour length $N$ when (a) the chain is confined to a spherical cavity, and (b) when it is placed in a viscoelastic fluid.

### A. Spherically confined polymers

For such polymers, the $N$-dependence of $t_c$ is found from Eq. (19), which contains two $N$-dependent factors: $N\langle R^2\rangle_{eq}$ and the integral $Z_1(x_0)$. The latter can actually be evaluated in closed form, as described in Appendix B. The result, when substituted into Eq. (19a), yields

$$t_c = \frac{\zeta N \langle R^2 \rangle_{eq}^{3/2}}{24\sqrt{2} k_B T a} \left[ 1 - \frac{a}{\sqrt{2} \langle R^2 \rangle_{eq}^{1/2}} \left\{ 2 - 2\ln 2 - \ln\left(\frac{1 - \sqrt{2} a / \langle R^2 \rangle_{eq}^{1/2}}{1 + \sqrt{2} a / \langle R^2 \rangle_{eq}^{1/2}}\right) + \ln\left(1 - \frac{2a^2}{\langle R^2 \rangle_{eq}}\right) \right\} \right]$$

(25)

This result exactly reproduces the scaling structure obtained by Doi[6] for the harmonic spring model. But as noted by Doi himself, the conclusions drawn from this expression should be treated with caution, since they are based on an approximation in which the true dynamics of the polymer is represented by the dynamics of its first mode. While this is not necessarily a poor approximation when considering such bulk properties as the viscosity, its application to chemical reactions can be more problematic, for the following reason. The overall behavior of $t_c$ is determined principally by the dynamics of the chain end-to-end vector, $\mathbf{R}(t)$, through the correlation function $\varphi(t)$. $\mathbf{R}(t)$ itself is principally determined by the first Rouse mode, $\mathbf{X}_1(t)$, but there are fluctuations around this value arising from the dynamics of the higher modes $\mathbf{X}_3(t)$, $\mathbf{X}_5(t)$, etc. As shown by Doi, their mean square amplitude, $A^2$, is roughly

$$A^2 = \langle (\mathbf{R}(t) - \mathbf{X}_1(t))^2 \rangle \approx 1.2 \langle R^2 \rangle_{eq} \sim N \tag{26}$$

so $A \sim \langle R^2 \rangle_{eq}^{1/2} \sim N^{1/2}$. The fluctuations in $\mathbf{R}(t)$ can therefore be said to occur in a sphere of radius $A$ centered on $\mathbf{X}_1(t)$, and because motion of $\mathbf{R}(t)$ in this sphere is very fast, the cyclization reaction is expected to occur whenever $|\mathbf{R}(t)| < A$, suggesting that we should actually use $A$ rather than $a$ in the expression for $t_c$ [Eq. (25)]. When this is done, we find that the $N$ dependence of $t_c$ is now determined principally by

$$t_c \sim N \langle R^2 \rangle_{eq} \qquad (27)$$

which has the following two limiting behaviors:

$$t_c \sim N^2, \qquad D \gg N \qquad (28a)$$

and

$$t_c \sim ND^2, \qquad N \gg D \qquad (28b)$$

The first of these scaling results [Eq. (28a)] is consistent with known behavior in free space[6,7,10,13] but the second is a new prediction that we believe it would be interesting to test experimentally.

In arriving at these results, we have had to appeal to physical arguments to justify the replacement of the reaction radius $a$ by $A$; however, a mathematical basis for this replacement can be suggested. Suppose the parameter $\sigma$ in Eq. (19b) is sufficiently small that the integrand there can be binomially expanded; then to leading order

$$Z_1 \approx 3\sigma/2 \sim \text{constant} \qquad (29)$$

Recalling the definition of $\sigma$ as $1 - 4x_0/3$, we see that the requirement that $\sigma$ be small is satisfied only if $x_0 \equiv 3a^2/2\langle R^2 \rangle_{eq}$ is of order 1, which means $a \sim \langle R^2 \rangle_{eq}^{1/2}$, exactly the condition derived by Doi.

**B. Polymers in viscoelastic media**

The *N*-dependence of $t_c$ for polymers in these conditions is now contained in Eqs. (24a) and (24b), but the integral $Z_2(x_0)$ can no longer be evaluated in closed form, and must be found numerically. However, if one were to treat the chain dynamics with the same "harmonic spring" approximation used earlier (i.e., retain only the lowest mode in the expression for the correlation function $\varphi(t)$), and then similarly assume that $\sigma$ takes on a small constant value, so that the integrand in $Z_2(x_0)$ can be binomially expanded to leading order, then $t_c$ is given by

$$t_c = BZ_3 N^{2/(2-2H)} \tag{30a}$$

where $B \equiv (3\sigma/(4-4H))(\zeta\Gamma(2H+1)/k\pi^2)^{1/(2-2H)}$ and $Z_3 \equiv \int_0^\infty dy\, y^{(2H-1)/(2-2H)} \varphi^2(y)$, with $\varphi(y) = E_{2-2H}(-y)$. In this approximation, therefore,

$$t_c \sim N^{2/(2-2H)}, \tag{30b}$$

and so $t_c \sim N^2$ when $H = 1/2$, and $t_c \sim N^4$ when $H = 3/4$. For these special values of *H*, the Mittag-Leffler function $E_{2-2H}(\cdots)$ reduces to simpler special functions (an exponential in the case $H = 1/2$, and an error function in the case $H = 3/4$.) This makes it a simple matter to evaluate Eq. (24b) [i.e., the integral $Z_2(x_0)$] essentially *exactly* by numerical methods, and the numerical results confirm the scaling predictions of Eq. (30b). We believe that Eq. (30b) holds for other values of *H* as well, although this remains to be confirmed. So if $\alpha_H$ denotes the exponent $2/(2-2H)$ in Eq. (30b), we can set down the following illustrative list of exponents:

$$\alpha_{0.5} = 2, \quad \alpha_{0.6} = 2.5, \quad \alpha_{0.65} = 2.9, \quad \alpha_{0.7} = 3.35, \quad \alpha_{0.75} = 4 \tag{31}$$

These results suggest that as $H$ becomes larger and chain relaxation becomes slower (essentially because of increased viscoelasticity), the chain cyclization rate becomes slower too, as seems reasonable. But the exponents in Eq. (31), for $H \geq 0.6$, are unexpectedly large. Most experimental studies of loop formation in polymers (usually carried out in dilute solution) have found exponent values in the range 2 (corresponding to chains at the theta point) to about 2.4 (corresponding to chains in good solvents.) Interestingly, a recent study (carried out at high viscosity under diffusion-limited conditions) of short (11 – 26 bases) segments of unstructured single-stranded DNA has found evidence that the cyclization time scales as the 3.85 power of the chain length. This unusually high exponent value has been attributed to electrostatic repulsion between the charged ends of the DNA.[38]

Further experimental and theoretical work on cyclization dynamics under conditions of confinement and crowding would clearly be helpful in clarifying some of the issues raised by the present calculations.

# APPENDIX A: PROPAGATOR OF A SPHERICALLY CONFINED POLYMER

The propagator $G(\mathbf{R},N)$ of Eq. (17) is the solution of Eq. (16) satisfying the boundary condition $G(|\mathbf{R}|=D,N)=0$. The solution is found by first transforming Eq. (16) to the spherical polar coordinates $R, \theta$ and $\phi$; when $N \neq 0$ and $|\mathbf{R}| \neq \mathbf{0}$, this yields

$$\left[\frac{\partial^2}{\partial R^2}+\frac{2}{R}\frac{\partial}{\partial R}+\frac{1}{R^2 \sin\theta}\frac{\partial}{\partial \theta}\sin\theta\frac{\partial}{\partial \theta}+\frac{1}{R^2 \sin^2\theta}\frac{\partial^2}{\partial \phi^2}\right]G = \frac{6}{b^2}\frac{\partial G}{\partial N} \quad (A.1)$$

which can be converted to the equation

$$\left[\frac{\partial^2}{\partial R^2}+\frac{1}{R}\frac{\partial}{\partial R}-\frac{1}{4R^2}+\frac{1}{R^2}\frac{\partial}{\partial \mu}(1-\mu^2)\frac{\partial}{\partial \mu}+\frac{1}{R^2(1-\mu^2)}\frac{\partial^2}{\partial \phi^2}\right]F = \frac{6}{b^2}\frac{\partial F}{\partial N} \quad (A.2)$$

by introducing the change of variables $\mu = \cos\theta$ and $F = \sqrt{R}\,G$. In this form, Eq. (A.2) is readily solved by writing $F$ in terms of variable-separated functions, i.e., as $F(R,\theta,\phi,N) = f(N)\Psi(R)M(\mu)\Phi(\phi)$. The substitution of this proposed solution into Eq. (A.2) leads to equations for $f, \Psi, M$ and $\Phi$ that are either trivially solvable or that can be recognized as the differential equations of known special functions. A general solution of Eq. (A.2) (and from there of Eq. (A.1)) is obtained by linearly combining the solutions involving $f, \Psi, M$ and $\Phi$. In this way, we find, after some algebra, that[39]

$$G(R,\mu,\phi,N) = \frac{1}{\sqrt{2\pi}}R^{-1/2}\sum_{l,m,n}A_{lmn}J_{l+1/2}(y_{l,n}R/D)C_{lm}P_l^m(\mu)e^{im\phi}\exp(-\lambda_{lmn}^2 Nb^2/6) \quad (A.3)$$

where $A_{lmn}$ is an as yet unknown expansion coefficient, $\lambda_{lmn}^2$, $m^2$ and $l(l+1)$ are constants of separation, with $m = 0, \pm 1, \pm 2, \ldots$ and $l = 0, 1, 2, \ldots$, $C_{lm} = \sqrt{(2l+1)(l-m)!/2(l+m)!}$, $J_{l+1/2}(\cdots)$ is a Bessel function of order $l+1/2$, $y_{l,n}$ is the *n*th zero of $J_{l+1/2}(\cdots)$ (i.e., $J_{l+1/2}(y_{l,n}) = 0$), and $P_l^m(\cdots)$ is an associated Legendre polynomial.

To determine the explicit form of the parameters $\lambda_{lmn}$, we substitute Eq. (A.3) into Eq. (A.1) (after introducing the variable change $\mu = \cos\theta$), and simplify the resulting expression by using the differential equations satisfied by the Bessel and associated Legendre functions. These steps lead to the identification

$$\lambda_{lmn}^2 = y_{l,n}^2 / D^2 \tag{A.4}$$

The parameters $A_{lmn}$ are determined by requiring that the propagator $G$ reduce to the function $R^{-2}\delta(R - R_0)\delta(\phi - \phi_0)\delta(\mu - \mu_0)$ when $N = 0$, with $R_0, \phi_0$ and $\mu_0$ some arbitrary initial values (which subsequently will be set to 0, 0, 0.) This requirement leads to

$$A_{lmn} = \frac{1}{D^2 J_{l+3/2}^2(y_{l,n})}\sqrt{\frac{2}{\pi R_0}} C_{lm} J_{l+1/2}(y_{l,n} R_0 / D) P_l^m(\mu_0) e^{-im\phi_0} \tag{A.5}$$

so the complete expression for the propagator is

$$G(\mathbf{R}, N \mid \mathbf{R}_0) = \frac{1}{\pi D^2 \sqrt{RR_0}} \sum_{l,m,n} \frac{C_{lm}^2}{J_{l+3/2}^2(y_{l,n})} J_{l+1/2}(y_{l,n} R / D) J_{l+1/2}(y_{l,n} R_0 / D) \times$$

$$\times P_l^m(\mu) P_l^m(\mu_0) e^{im(\phi - \phi_0)} \exp\left(-Nb^2 y_{l,n}^2 / 6D^2\right) \tag{A.6}$$

To pass to the limit $\mathbf{R}_0 \to \mathbf{0}$ in this expression, one separates the $l = 0$ contribution to the sum from the remaining terms, substitutes the general Bessel relation $J_\nu(z) = (z/2)^\nu \sum_{k=0}^{\infty} (-z/2)^{2k} / k!\Gamma(\nu + k + 1)$ into the result, and then sets $R_0 = 0$, thus producing Eq. (17).

# APPENDIX B. EVALUATION OF $Z_1(x_0)$ [Eq. (19a)]

To evaluate $Z_1(x_0)$ in closed form, we first evaluate the indefinite integral $Z_1^{(I)}(x_0) \equiv \int dy\, y^{-1}[(1-\sigma y)^{-3/2} - 1]$, and then find $Z_1(x_0)$ from

$$Z_1(x_0) = \lim_{\varepsilon \to 0} Z_1^{(I)}(x_0)\Big|_\varepsilon^1 \qquad (B.1)$$

One may verify by differentiation that $Z_1^{(I)}(x_0)$ is given by

$$Z_1^{(I)}(x_0) = \frac{2}{\sqrt{1-\sigma y}} + \ln\left(1 - \sqrt{1-\sigma y}\right) - \ln\left(1 + \sqrt{1-\sigma y}\right) - \ln y \qquad (B.2)$$

Hence,

$$Z_1^{(I)}(x_0)\Big|_\varepsilon^1 = 2\left(\frac{1}{\sqrt{1-\sigma}} - \frac{1}{\sqrt{1-\sigma\varepsilon}}\right) + \ln\left(\frac{1-\sqrt{1-\sigma}}{1+\sqrt{1-\sigma}}\right) - \ln\left(\frac{1-\sqrt{1-\sigma\varepsilon}}{1+\sqrt{1-\sigma\varepsilon}}\right) + \ln\varepsilon \qquad (B.3)$$

In the limit $\varepsilon \to 0$, this expression becomes

$$Z_1^{(I)}(x_0)\Big|_\varepsilon^1 = 2\left(\frac{1}{\sqrt{1-\sigma}} - 1\right) + \ln\left(\frac{1-\sqrt{1-\sigma}}{1+\sqrt{1-\sigma}}\right) - \ln\varepsilon - \ln(\sigma/2) + \ln 2 + \ln\varepsilon + O(\varepsilon) \qquad (B.4)$$

which, after substitution in Eq. (B.1), leads to

$$Z_1(x_0) = \frac{2}{\sqrt{1-\sigma}} - 2 + 2\ln 2 + \ln\left(\frac{1-\sqrt{1-\sigma}}{1+\sqrt{1-\sigma}}\right) - \ln\sigma \qquad (B.5)$$

which in turn produces

$$Z_1(x_0) = \frac{\sqrt{2}\langle R^2\rangle_{eq}^{1/2}}{a} - 2 + 2\ln 2 + \ln\left(\frac{1-\sqrt{2}\,a/\langle R^2\rangle_{eq}^{1/2}}{1+\sqrt{2}\,a/\langle R^2\rangle_{eq}^{1/2}}\right) - \ln\left(1 - 2a^2/\langle R^2\rangle_{eq}\right) \qquad (B.6)$$

after putting in the definition of $\sigma$.

[16] I. M. Sokolov, Phys. Rev. Lett. **90**, 080601 (2003).

[17] P. Debnath and B. J. Cherayil, J. Chem. Phys. **120**, 2482 (2004).

[18] J. Z. Y. Chen, H.-K. Tsao and Y.-J. Sheng, Phys. Rev. E 72, 031804 (2005); N. M. Toan, G. Morrison, C. Hyeon and D. Thirumalai, J. Phys. Chem. B **112**, 6094 (2008).

[19] R. R. Cheng, T. Uzawa, K. W. Plaxco and D. E. Makarov, J. Phys. Chem. B **113**, 14026 (2009).

[20] R. Shusterman, S. Alon, T. Gavrinyov and O. Krichevsky, Phys. Rev. Lett. **92**, 048303 (2004); M. Weiss, M. Elsner, F. Kartberg and T. Nilsson, Biophys. J. **87**, 3518 (2004); D. S. Banks and C. Fradin, *ibid*. **89**, 2960 (2005); I. Bronstein, Y. Israel, E. Kepten, S. Mai, Y. Shav-Tal, E. Barkai and Y. Garini, Phys. Rev. Lett. **103**, 018102 (2009); J. T. Mika and B. Poolman, Curr. Opinion Biotech. **22**, 117 (2011).

[21] S. Yang and J. Cao, J. Chem. Phys. **121**, 572 (2004); C. Yeung and B. Friedman, J. Chem. Phys. **122**, 214909 (2005).

[22] S. C. Chaturvedi, in *Stochastic Processes*: *Formalism and Applications, Lecture Notes in Physics*, Vol. 184, ed. by G. S. Agarwal and S. Dattagupta (Springer-Verlag, Berlin, 1983)

[23] The angular integrations are carried out by noting that the dot product $\mathbf{R} \cdot \mathbf{R}_0$ in Eq. (4) can be written as $RR_0 \cos\psi$, where $\psi$, the angle between the vectors $\mathbf{R}$ and $\mathbf{R}_0$, is a function of the angular variables $\theta, \phi, \theta_0$ and $\phi_0$. The integral over these variables involves the function $e^{h\psi}$, $h$ standing for some arbitrary coefficient. This function itself can be rewritten identically as $e^{h\psi} = 4\pi\sqrt{\pi/2h} \sum_{l=0}^{\infty} \sum_{m=-l}^{l} I_{l+1/2}(h) Y_{l,m}(\theta,\phi) Y_{l,m}^{*}(\theta_0,\phi_0)$, where $I_\nu(z)$ is a modified Bessel function of order $\nu$, and $Y_{l,m}(\theta,\phi)$ is a spherical
25

harmonic. Application of the formula $\int_0^\pi d\theta \sin\theta \int_0^{2\pi} d\phi\, Y_{l,m}(\theta,\phi) = \sqrt{4\pi}\,\delta_{m,0}\delta_{l,0}$ then selects out a single term corresponding to $l = 0$ in the above double sum, and substitution of the result $I_{1/2}(z) = \sqrt{2/\pi z}\,\sinh z$ then finally leads to Eq. (5).

**FIGURE CAPTIONS**

1. The normalized end-to-end distance correlation function $\varphi(t)$ as a function of the dimensionless time $tk_B T/b^2\zeta$, as given by Eq. (23), at the chain length $N=100$ and the dimensionless reaction radius $a/b=1.0$, for the following values of $H$: 0.5 (blue), 0.6 (black), 0.65 (red), 0.7 (magenta) and 0.75 (green). .

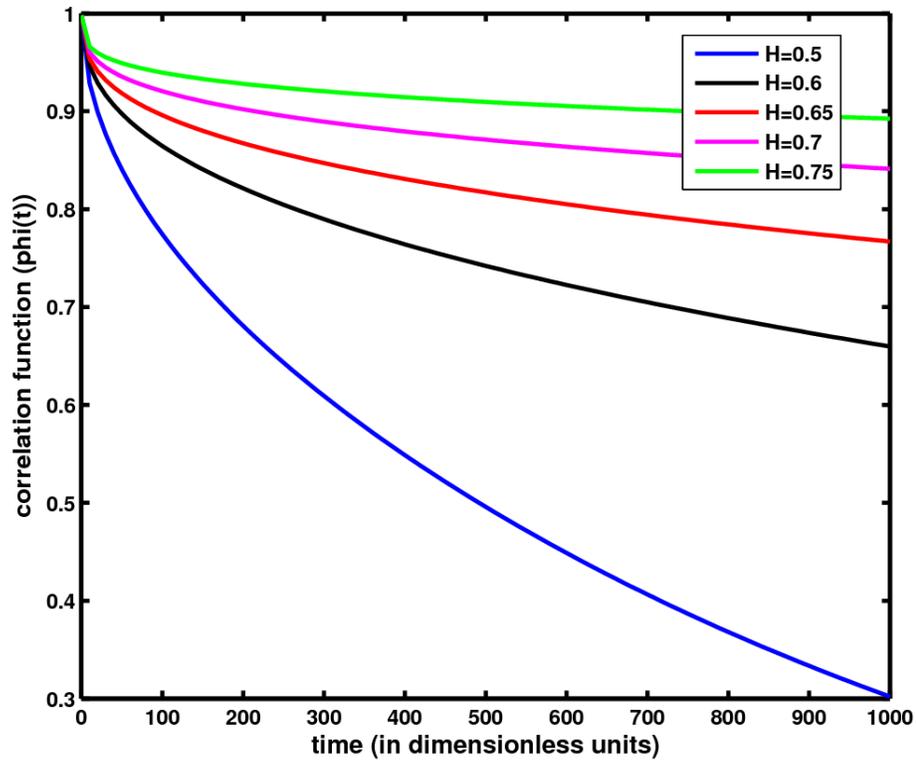

**FIGURE 1**